\documentclass[twocolumn,english,10pt,aps,prd,a4paper,preprintnumbers,floatfix,nofootinbib,showpacs,superscriptaddress, notitlepage]{revtex4-1} 
\pdfoutput=1
\usepackage{graphicx}
\usepackage{amsmath}
\usepackage{amssymb}
\usepackage{amsfonts}
\usepackage{physics}
\usepackage[dvipsnames]{xcolor}
\usepackage[linktoc=none]{hyperref}

\usepackage{braket}
\usepackage{tabstackengine}
\stackMath
\usepackage[mathscr]{euscript}

\hypersetup{colorlinks,linkcolor={blue},citecolor={teal},urlcolor={violet}}

\newcommand{\INFN}{INFN - Sezione di Napoli, Complesso Univ. Monte S. Angelo, I-80126 Napoli, Italy}
\newcommand{\UNINA}{Dipartimento di Fisica ``Ettore Pancini'', Università degli studi di Napoli ``Federico II'', Complesso Univ. Monte S. Angelo, I-80126 Napoli, Italy}
\newcommand{\SSM}{Scuola Superiore Meridionale, Largo San Marcellino 10, 80138 Napoli, Italy}

\newcommand{\MPBH}{{M}_{\rm PBH}}

\newcommand{\GNt}{\Gamma_{22}}

\newcommand{\varrhopbh}{\varrho_{\rm PBH}}
\newcommand{\varrhorad}{\varrho_{\rm rad}}

\newcommand{\Mpl}{{M}_{\rm Pl}}

\newcommand{\GeV}{{\rm GeV}}

\newcommand{\Lagr}{\mathcal{L}}
\newcommand{\Tsphal}{T_{\rm sphal}}
\newcommand{\Tewpt}{T_{\rm EWPT}}

\begin{document}

\title{Impact of primordial black holes on heavy neutral leptons searches\\in the framework of resonant leptogenesis}

\author{Roberta Calabrese}
\email{rcalabrese@na.infn.it}
\affiliation{\UNINA}
\affiliation{\INFN}

\author{Marco Chianese}
\email{marco.chianese@unina.it}
\affiliation{\UNINA}
\affiliation{\INFN}

\author{Jacob Gunn}
\email{jacobwilliam.gunn@unina.it}
\affiliation{\UNINA}
\affiliation{\INFN}

\author{Gennaro Miele}
\email{miele@na.infn.it}
\affiliation{\UNINA}
\affiliation{\INFN}
\affiliation{\SSM}

\author{Stefano Morisi}
\email{smorisi@na.infn.it}
\affiliation{\UNINA}
\affiliation{\INFN}

\author{Ninetta Saviano}
\email{nsaviano@na.infn.it}
\affiliation{\INFN}
\affiliation{\SSM}

\begin{abstract}
We investigate the effects on sub-TeV resonant leptogenesis of Primordial Black Holes with masses from $10^6$ to $\sim 10^9$~g. The latter might dominate the energy content of the Universe altering its evolution and, eventually, diluting the final baryon asymmetry. We find that, depending on the mass and abundance of Primordial Black Holes, the parameter space of sub-TeV resonant leptogenesis shrinks towards higher Right-Handed Neutrino masses and smaller active-sterile mixing. Remarkably, this translates into important implications for the experimental searches of heavy neutral leptons. Conversely, we demonstrate that a possible future detection of sub-TeV heavy neutral leptons would disfavour regions of the parameter space of Primordial Black Holes currently allowed.
\end{abstract}

\maketitle

\section{Introduction \label{sec:intro}}

Among the most attractive theoretical models proposed to explain the presently very small value of B, the so-called cosmological Baryon Asymmetry in the Universe (BAU), the scenario of baryogenesis trough leptogenesis was proposed by Fukugita and Yanagida~\cite{Fukugita:1986hr}. According to this model, $L$ and CP violating decays of singlet neutrinos $N_i$ with considerably large Majorana masses, produce initially an excess in the lepton number $L$ which is then converted into the observed $B$ asymmetry through $(B + L)$-violating sphaleron interactions. This simple and elegant built-in generation is intimately linked to the seesaw mechanism, whose best feature is precisely the existence of Right-Handed Neutrinos (RHNs) $N_i$ invoked to explain the origin of neutrino masses, and the mass hierarchy between neutrinos and charged leptons. Within this mechanism RHNs are expected to have mass close to the Grand Unified Theory (GUT) scale. However, this is just our prejudice and in principle sterile neutrinos can have any mass. This is true also in relation to leptogenesis. Indeed, in the framework of high-scale vanilla leptogenesis one typically has $M_1 \ll M_2\ll M_3 $, each not too far from the GUT energy scale $M_{\rm GUT}\sim 10^{16}$~GeV. In this case, in order to reproduce the observed baryon asymmetry it is necessary that $M_1\gtrsim 10^9$GeV~\cite{Davidson:2002qv}. However, in case of degenerate RHNs masses $M\sim M_1\sim M_2$ ($M_3$ can be at the GUT scale), it is possible to escape this limit. Indeed, it has been shown that the right baryon asymmetry can be achieved with $M\sim$~TeV by means of resonant leptogenesis~\cite{Pilaftsis:1997jf,Pilaftsis_2004}, where the leptonic CP asymmetry can be resonantly enhanced if the RHNs mass differences is comparable to their decay widths. More recently, many studies have pointed out that successful resonant leptogenesis can be extended to sub-TeV scales, down to $\mathcal{O}(1-100)$~GeV~\cite{Canetti:2012kh,BhupalDev:2014pfm,BhupalDev:2014oar,Hambye:2016sby,Bambhaniya:2016rbb,Granelli:2020ysj,Klaric:2020phc}. From the theoretical point of view, a lot of effort has been put into the so-called low-energy seesaw mechanisms like type-I seesaw~\cite{Asaka:2005an} and inverse-seesaw~\cite{Mohapatra:1986bd} (for a review see Ref.~\cite{Boucenna:2014zba}). From an experimental side, current and planned laboratory facilities are constraining the existence of heavy neutral leptons with masses around GeV scale through their mixing with active neutrinos~\cite{Bauer:2019vqk, MATHUSLA:2022sze, Baldini:2021hfw,Feng:2022inv,Abdullahi:2022jlv}. Remarkably, the baryogenesis via resonant leptogenesis provides important information on the parameter space of heavy neutral leptons searches.

Different leptogenesis scenarios, and the relative parameters space, can be also constrained by the possible presence in the early Universe of Primordial Black Holes (PBHs), hypothetical black holes possibly formed after the inflationary epoch~\cite{Zeldovich:1967lct, Hawking:1971ei, Carr:1974nx, Hawking:1974rv, Carr:1976zz,Escriva:2021aeh, Escriva:2022duf, Ozsoy:2023ryl}. In recent years, the interest in PBHs has been significantly renewed thanks to their several interesting consequences in astrophysics and cosmology~\cite{Green:2020jor, Carr:2020gox, Carr:2021bzv}. In particular, by emitting Hawking radiation on observable timescales, PBHs could evaporate to all elementary particles, regardless of their Standard Model (SM) gauge interactions, leading for example to a nonthermal production of heavy neutral leptons, namely RHNs. Moreover, PBHs can dominate the Universe in early times, drastically changing its evolution and leading to a matter-dominated epoch. Both these effects have been demonstrated to have significant implications for different leptogenesis scenarios~\cite{Fujita:2014hha, Hamada:2016jnq, Morrison:2018xla,Perez-Gonzalez:2020vnz, Datta:2020bht, JyotiDas:2021shi, Bernal:2022pue, Calabrese:2023key, Schmitz:2023pfy}. Following these recent studies, in this  work we focus on the effect of PBHs with masses from $10^6$ to $\sim10^9$~g on the predictions of sub-TeV resonant leptogenesis,  being interested in low-scale neutrino mass generation. In particular, we compute the final baryon asymmetry by solving the relevant set of coupled Boltzmann equations related to leptogenesis together with the PBHs evolution, and we show how the presence of PBHs can strongly modify the parameter space of heavy neutral leptons searches.

The paper is organized as follows. In Sec.~\ref{sec:leptogenesis} we discuss the scenario of sub-TeV resonant leptogenesis. In particular, we describe the seesaw model and the relevant Boltzmann equations to compute the final baryon asymmetry of the Universe, and we report the standard constraints on the active-sterile mixing parameter as a function of the right-handed neutrino mass scale. In Sec.~\ref{sec:PBH} we delineate the physics of PBHs and their impact on the evolution of cosmological observable. Then, in Sec.~\ref{sec:results} we report our main results which quantify the interplay between the parameter space of sub-TeV resonant leptogenesis and the one of PBHs. Finally, in Sec.~\ref{sec:conclusions} we draw our conclusions.

\section{Sub-TeV resonant leptogenesis \label{sec:leptogenesis}}

In this section we overview the mechanism of sub-TeV resonant leptogenesis describing the main parameters (connected to the seesaw mechanism for the neutrino mass and mixing generations) and the relevant Boltzmann equations. Moreover, we discuss the parameter space achieving the correct baryon asymmetry of the Universe.

\subsection*{Type-I seesaw}

One of the simplest mechanisms to generate neutrino masses is the type-I seesaw, in which the SM is minimally extended by $n$ Majorana singlet fermions, the Right-Handed Neutrinos (RHNs), $N_i$ with $i=1,...,n$, which mix with the SM leptons through their Yukawa couplings to the Higgs. The simplest case is $n=2$ predicting the lightest active neutrino to be massless. However, ${\rm SO}(10)$ models predict $n=3$, and in general having the same number of left and right handed neutrino fields is a quite common assumption. Here, we assume $n=3$ and consider two nearly degenerate RHNs ($M\sim M_1 \sim M_2$) with a tiny mass splitting $\Delta M = M_2 - M_1 \ll M_{1,2}$ and a third effectively decoupled RHN ($M_3 \gg M$). We note that due to radiative corrections the mass splitting is expected to be at the level of the active neutrino mass splitting, i.e. in the range ($10^{-13} - 10^{-11})$ GeV. However, it could be higher or with some fine-tuning can be smaller~\cite{Roy:2010xq}.

The Lagrangian reads simply
\begin{equation}
    \Lagr_{\rm seesaw} =  \frac{1}{2}\bar{N}^c_i \hat{M}_{ij} N_j - Y_{\ell i}\bar{L}_\ell \tilde{\phi} N_i + {\rm h.c.}, 
\end{equation}
where $\hat{M}$ is the mass matrix of the RHNs (we work in the basis where this is diagonal), and $Y_{\ell i}$ is the Yukawa matrix element coupling the SM leptons of flavour $\ell = e,\mu,\tau$ to $N_i$, through the Higgs doublet $\tilde{\phi} = i\sigma_2 \phi^*$. 

The Yukawa matrix is usually expressed in a most convenient way through the Casas-Ibarra parameterisation~\cite{Casas:2001sr}
\begin{equation}\label{Y}
Y = \frac{1}{v_0}U_{\rm PMNS}\cdot \sqrt{\hat{m}_{\nu}}\cdot R^T \cdot \sqrt{\hat{M}} \,,
\end{equation}
where $v_0=246~\GeV$ is the Higgs vacuum expectation value at zero temperature, $U_{\rm PMNS}$ is the Pontecorvo-Maki-Nakagawa-Sakata matrix describing the mixing in the active neutrino sector, $\hat{m}_{\nu} = \mathrm{diag}(m_1,m_2,m_3)$ is the diagonal active neutrino mass matrix, and $R$ is a complex rotation matrix parameterised by three complex angles, $z_{12},z_{13},z_{23}$. For the mass and mixing neutrino parameters, we take the best-fit values from Ref.~\cite{Capozzi:2021fjo} (see also Refs.~\cite{Esteban:2020cvm, deSalas:2020pgw}). Moreover, we assume the two neutrino Majorana phases to be zero and  we consider the normal neutrino mass ordering ($m_1<m_2<m_3$) with $m_1=0$, $m_2 = \sqrt{m_{\rm sol}^2}$ and $m_3 = \sqrt{m_{\rm atm}^2}$. In this case, which is equivalent to the one discussed in Ref.~\cite{Granelli:2020ysj}, the matrix $R$ has only one physically relevant angle $z_{23}=x+i\,y$, with real $x,y$. We therefore have
\begin{equation}\label{eq:R}
R = \begin{pmatrix}
0 & \cos (z_{23})  & \sin (z_{23})  \\
0 & -\sin (z_{23}) & \cos  (z_{23})  \\
1 & 0 & 0 \\
\end{pmatrix} \,.
\end{equation}
In this framework, the mixing parameter between the active light neutrinos and heavy sterile ones takes the following expression~\cite{Chianese:2018agp}
\begin{eqnarray}
    U^2 & = & \sum_{\alpha\,N} |U_{\alpha (N+3)}|^2 \\
    & = & \frac{m_2-m_3}{2} \frac{\Delta M}{M^2}\cos(2 x) + \frac{(m_2+m_3)}{M} \cosh(2y) \nonumber \,,
    \label{eq:U2}
\end{eqnarray}
where the contribution from $N_3$ is negligible. This is the most relevant quantity that one aims to experimentally constrain. Indeed, there already exist several theoretical and experimental limits, and forecast bounds in the plane $M$-$U^2$ from numerous upcoming, planned and proposed experiments~\cite{Bauer:2019vqk,MATHUSLA:2022sze, Baldini:2021hfw,Feng:2022inv,Abdullahi:2022jlv}. Among the theoretical bounds, the requirement of successful baryogenesis via resonant leptogenesis provides specific allowed regions in the plane $M$-$U^2$, see for instance Ref.~\cite{Klaric:2020phc}.

\subsection*{Boltzmann equations}

In order to determine the comoving number density of the baryon asymmetry $\mathcal{N}_B$, we track the evolution of the comoving number densities $\mathcal{N}_{\Delta \ell}$ and $\mathcal{N}_{N_i}$ of the leptonic asymmetry $\Delta \ell$ in the flavour $\ell$ and of $N_i$, respectively. The comoving number densities are simply equal to $\mathcal{N} = n\,a^3$ where $n$ is the physical number density and $a$ the scale factor. In our framework, only $N_1$ and $N_2$ contribute to the final asymmetry since $N_3$ is effectively decoupled. Moreover, we focus on the case where all the RHNs $N_i$ thermalise long before leptogenesis, which we refer to as thermal initial abundance. Under this assumption, the effects of RHNs oscillation ($N_i \leftrightarrow N_j$) has been shown to be subdominant~\cite{Klaric:2021cpi, Akhmedov:1998qx, Drewes:2016gmt}, thus ensuring the validity of the Boltzmann equation formalism. Hence, the evolution of $\mathcal{N}_{\Delta \ell}$ and $\mathcal{N}_{N_i}$ can be simply described by the two coupled Boltzmann equations which account for the changes in the number density due to the following processes~\cite{Hambye:2016sby,Giudice:2003jh}:
\begin{itemize}
    \item $1\leftrightarrow 2$ (inverse) decays of $N_{1,2}$ and the Higgs, $N_{1,2} \leftrightarrow \ell \phi^\dagger$ and $\phi \leftrightarrow N_{1,2} \ell$, with $\gamma_{D}$ denoting the corresponding reaction density. Due to the thermal corrections to the particle masses, the Higgs decay typically occurs at higher temperatures ($T \gg M$), producing the states $N_{1,2}$, while the decays $N_{i} \to \ell \phi^\dagger$ may become kinematically possible at lower temperatures ($T \sim M$) and deplete the population of $N_i$. At intermediate temperatures, neither decay mode is kinematically possible.
    \item $2 \leftrightarrow 2$ scatterings with $\Delta L = 1$, involving (top) quark or gauge boson final states mediated by leptons or Higgs, with reaction densities $\gamma_{S_s}$ and $\gamma_{S_t}$ for $s$-channel and $t$-channel processes, respectively. 
    \item $2 \leftrightarrow 2$ scatterings with $\Delta L = 2$, which are mediated by $N_{1,2}$. However, their contribution is negligible and therefore not considered here.
\end{itemize}
The Boltzmann equations in terms of $\alpha = \log_{10} a$ read~\cite{Hambye:2016sby,Giudice:2003jh}
\begin{eqnarray} \label{eq:BEN} \frac{{\rm d}\mathcal{N}_{N_i}}{{\rm d}\alpha} &=& \frac{a^3\ln(10)}{H}\left(1-\frac{\mathcal{N}_{N_i}}{\mathcal{N}_N^{\rm eq}}\right)(\gamma_{D} + 2\gamma_{S_s} + 4\gamma_{S_t}) \\
\label{eq:BEL}
\frac{{\rm d}\mathcal{N}_{\Delta \ell}}{{\rm d}\alpha} &=&\frac{a^3\ln(10)}{H} \sum_i \left[ \epsilon_{\ell \ell}^i \left(\frac{\mathcal{N}_{N_i}}{\mathcal{N}_N^{\rm eq}} - 1\right)\gamma_{D} \right. \\
\nonumber &&\left. -P_{\ell\,i}\frac{\mathcal{N}_{\Delta \ell}}{\mathcal{N}_\ell^{\rm eq}} \left(2\gamma_{D} + 2\gamma_{S_t} + \frac{\mathcal{N}_{N_i}}{\mathcal{N}_{N_i}^{\rm eq}} \gamma_{S_s}\right) \right] \,,
\end{eqnarray}
where $H$ is the Hubble rate and $\mathcal{N}_{X}^{\rm eq}$ is the comoving number density of species $X$ in thermal equilibrium with the photon bath. The reaction density $\gamma_{D}$ for the $1\leftrightarrow2$ processes reads as
\begin{eqnarray}
    \gamma_{D} &=&
    \frac{M^3}{\pi^2 z} K_1\left(z\right)\Gamma_{N_i \to \phi L} \Theta(M - M_L + M_\phi) \nonumber \\
    &+& \frac{2 M_H^2 M}{\pi^2 z}K_1\left(z_\phi\right)\Gamma_{\phi \to N_i L} \Theta( M_\phi - M_L + M)
\end{eqnarray}
where $z_\phi = M_\phi / T$, and the quantities $\Gamma_{N_i \to \phi L}$ and $\Gamma_{\phi \to N_i L}$ are the decay widths corrected by the Higgs and lepton thermal masses
\begin{eqnarray}
    \Gamma_{N_i \to \phi L} &=& \frac{M(Y^\dagger Y)_{ii}}{8\pi} \lambda^{\frac{1}{2}}(1,a_\phi,a_L)\\
    & & \times (1-a_\phi + a_L)\,\Theta (1-a_\phi - a_L)\,,\nonumber\\
    \Gamma_{\phi \to N_i L} &=& \frac{M_\phi(Y^\dagger Y)_{ii}} {8\pi a_\phi^2}\lambda^{\frac{1}{2}}(a_\phi,1,a_L)\\
    & & \times (a_\phi - a_L-1)\,\Theta(a_\phi - a_L-1)\nonumber\,,
\end{eqnarray}
where $\lambda(a,b,c) = (a-b-c)^2 - 4bc$, $a_X = (m_X/M)^2$, and the Heaviside functions reflect the kinematic restrictions. Hence, depending on the thermal masses of particles, only one of the RHN decay or the Higgs might be kinematically allowed. The expressions for the rates $\gamma_{S_s}$ and $\gamma_{S_t}$ of the $2 \leftrightarrow 2$ quark and gauge boson scatterings can be found in Ref.~\cite{Giudice:2003jh}.

We account for the thermal corrections to the Higgs, lepton, quark, and boson masses. In particular, the Higgs thermal mass reads~\cite{Cline:1993bd}
\begin{equation}
    M_\phi(T)^2 = m_\phi^2 \frac{v^2(T)}{v_0^2} + \delta_{M_\phi} \Theta(T - \Tewpt)\,,
\end{equation}
where $v^2(T) = v_0^2 (1-T^2/\Tewpt^2)\Theta\left(\Tewpt-T\right)$ is the temperature-dependent Higgs vacuum expectation value with $\Tewpt = 160~{\rm GeV}$ being the temperature of the electroweak phase transition (EWPT), $m_\phi = 125~{\rm GeV}$ is the zero-temperature Higgs mass, and
\begin{equation}
	\delta_{M_\phi} = \left(\frac{\lambda_\phi}{2} +\frac{3g_2^2 + g_1^2}{16}  + \frac{y_t^2}{4} \right)\left(T^2-\Tewpt^2\right)\,,
\end{equation}
where $\lambda_\phi$ is the Higgs quartic coupling, $g_2$ and $g_1$ are $SU(2)_L$ and $U(1)_Y$ couplings, and $y_t$ is the top-quark Yukawa coupling. Similar expressions hold for the thermal masses of the other SM particles~\cite{Giudice:2003jh}, while the thermal corrections to the RHNs masses are negligible due to the smallness of the Yukawa couplings.

The Boltzmann equation~\eqref{eq:BEL} for the flavored leptonic asymmetry $\Delta \ell$ depends on the flavoured CP asymmetry parameter $\epsilon^i_{\ell \ell}$ and on the projection probability $P_{\ell i}= |Y_{\ell i}|^2/(Y^\dagger Y)_{ii}$ of heavy neutrino mass state $i$ on to flavour state $\ell$. The flavoured CP asymmetry parameter is given by~\cite{Granelli:2020ysj} (see also Refs.~\cite{BhupalDev:2014pfm,BhupalDev:2014oar, Hambye:2016sby})
\begin{equation}
\label{epsilon}
\epsilon_{\ell\ell}^i = \sum_{j,\,j\neq i}\frac{2\, \mathrm{sgn}(M_i-M_j)\,I_{ij,\ell\ell}\,x^0\gamma(z)}{4\frac{\GNt}{\Gamma_{jj}}(x^0 + x_T(z))^2 + \frac{\Gamma_{jj}}{\GNt}\gamma^2(z)} \,,
\end{equation}
where $z = M/T$, and $x^0,\,x_T$ encode the zero temperature and thermally corrected mass splittings as $x^0 = \Delta M/\Gamma_{22}$ and
\begin{equation}
    x_T \simeq \frac{\pi}{4z^2}\sqrt{\left(1-\frac{\Gamma_{11}}{\GNt}\right)^2 + 4\frac{|\Gamma_{12}|^2}{\Gamma_{22}^2}} \,,
\end{equation}
with $\Gamma_{ij} = (Y^\dagger Y)_{ij}\sqrt{M_iM_j}/8\pi$, and $I_{ij,\ell\ell}$ is given by
\begin{equation}
\label{eq:I}
I_{ij,\ell\ell} =  \frac{\mathrm{Im}(Y^\dagger_{i\ell}Y_{\ell\,j}(Y^\dagger Y)_{ij} ) + \frac{M_i}{M_j}\mathrm{Im}(Y^\dagger_{i\ell}Y_{\ell\,j}(Y^\dagger Y)_{ji} ) }{(Y^\dagger Y)_{ii}(Y^\dagger Y)_{jj}} \,,
\end{equation}
and the function $\gamma(z)$ describes the temperature dependence of the RHN self-energy cut~\cite{Hambye:2016sby}, which we take from figure~1 in Ref.~\cite{Granelli:2020ysj} without considering the interpolation to account the collinear emissions of soft gauge bosons.

Differently from high-scale leptogenesis where the lepton asymmetry is frozen in long before the sphaleron decoupling at $\Tsphal \approx 132~{\rm GeV}$, in our scenario the lepton asymmetry evolves during and even after the sphaleron decoupling when $M < \Tsphal$. Therefore, to compute the final baryon asymmetry we solve the additional Boltzmann equation accounting for non-instantaneous sphaleron freeze-out~\cite{Eijima:2017cxr}:
\begin{equation}
\frac{{\rm d}\mathcal{N}_B}{{\rm d}\alpha} =   -\frac{\ln(10)}{H}\Gamma_B(T)( \mathcal{N}_B + \chi(T)\mathcal{N}_{\Delta} )\,,
\label{eq:freezeout}
\end{equation}
where $\mathcal{N}_{\Delta}  = \sum_{\ell} \mathcal{N}_{\Delta \ell}$ is the total lepton asymmetry,
\begin{eqnarray}
    \chi (T) = 4\frac{77 + 27(v(T)/T)^2}{869 + 333(v(T)/T)^2}\,,
\end{eqnarray}
and $\Gamma_B$ is the rate of the sphaleron processes. We have~\cite{Eijima:2017cxr}
\begin{equation}
\Gamma_B = 9 \frac{869 + 333(v(T)/T)^2}{792 + 306(v(T)/T)^2} \frac{\Gamma_{\rm CS}(T)}{T^3} \,,
\end{equation}
with $\Gamma_{\rm CS}$ being the  Chern-Simons diffusion rate
\begin{equation}
    \Gamma_{\rm CS}  = \begin{cases}
        T^4 \, {\rm exp}(-14.7 + 0.83\,T) & T < \Tewpt \\
        18 \,T^4\, a^5_W & T > \Tewpt
    \end{cases}\,,
\end{equation}
with $a_W \approx 0.0073$ being the fine structure constant. As long as $\Gamma_B > H$, the comoving baryon asymmetry is kept to be equal to $\mathcal{N}_B = -\chi(T) \mathcal{N}_\Delta$. In the approximation of instantaneous sphaleron freeze-out, one simply considers $\mathcal{N}_B \approx -\chi(\Tsphal) \mathcal{N}_\Delta$ where $\Tsphal$ is obtained from the condition $\Gamma_B(\Tsphal)=H(\Tsphal)$. The usual efficiency factors $\chi = 28/79$ and $\chi = 12/37$ for the symmetric and broken phases are recovered for $\Tsphal \gg \Tewpt$ and $\Tsphal \ll \Tewpt$, respectively. However, this is not a good approximation in the case under study since $\mathcal{N}_\Delta$ is still evolving when $\Gamma_B \approx H$.

\subsection*{The parameter space of sub-TeV resonant leptogenesis}

We solve the three Boltzmann equations~\eqref{eq:BEN},~\eqref{eq:BEL} and~\eqref{eq:freezeout} assuming $\mathcal{N}_{N_i} = \mathcal{N}^{\rm eq}_{N_i}$ (thermal abundance), $\mathcal{N}_{\Delta \ell} = 0$ and $\mathcal{N}_{B} = 0$ as initial conditions and compute the final baryon asymmetry yield $Y_B = \mathcal{N}_B/\mathcal{S}$ with $\mathcal{S}$ being the comoving entropy density. This has to be compared with the observed value as measured by the Planck experiment~\cite{Planck:2018yye}
\begin{equation}
    Y^{\rm obs}_B = (8.75\pm0.23)\times 10^{-11}\,.
\end{equation}
In the model we consider for sub-TeV resonant leptogenesis, the final yield $Y_B$ depends on four main parameters: the RHN mass scale $M\sim M_1 \sim M_2$, the tiny mass splitting $\Delta M = M_2 - M_1$, and the two angles $x$ and $y$ of the Casas-Ibarra matrix. However, its dependence on $x$ is periodic. In the present analysis, we fix $x=3\pi/4$ which maximises the final baryon asymmetry and, as we will see later, will provide conservative constraints on PBHs whose main effect is the reduction of the asymmetry due to entropy injection (see next section). Hence, we focus on investigating the dependence of $Y_B$ on the remaining three parameters $M$, $\Delta M$ and $y$, which together determine $U^2$ (see  Eq.~\eqref{eq:U2}). This allows us to show the regions achieving successful leptogenesis ($Y_B = Y_B^{\rm obs}$) in the plane $M$-$U^2$ for different values of relative mass splitting $\Delta M/M$.
\begin{figure}[t!]
    \centering
    \includegraphics[width = 0.9 \linewidth]{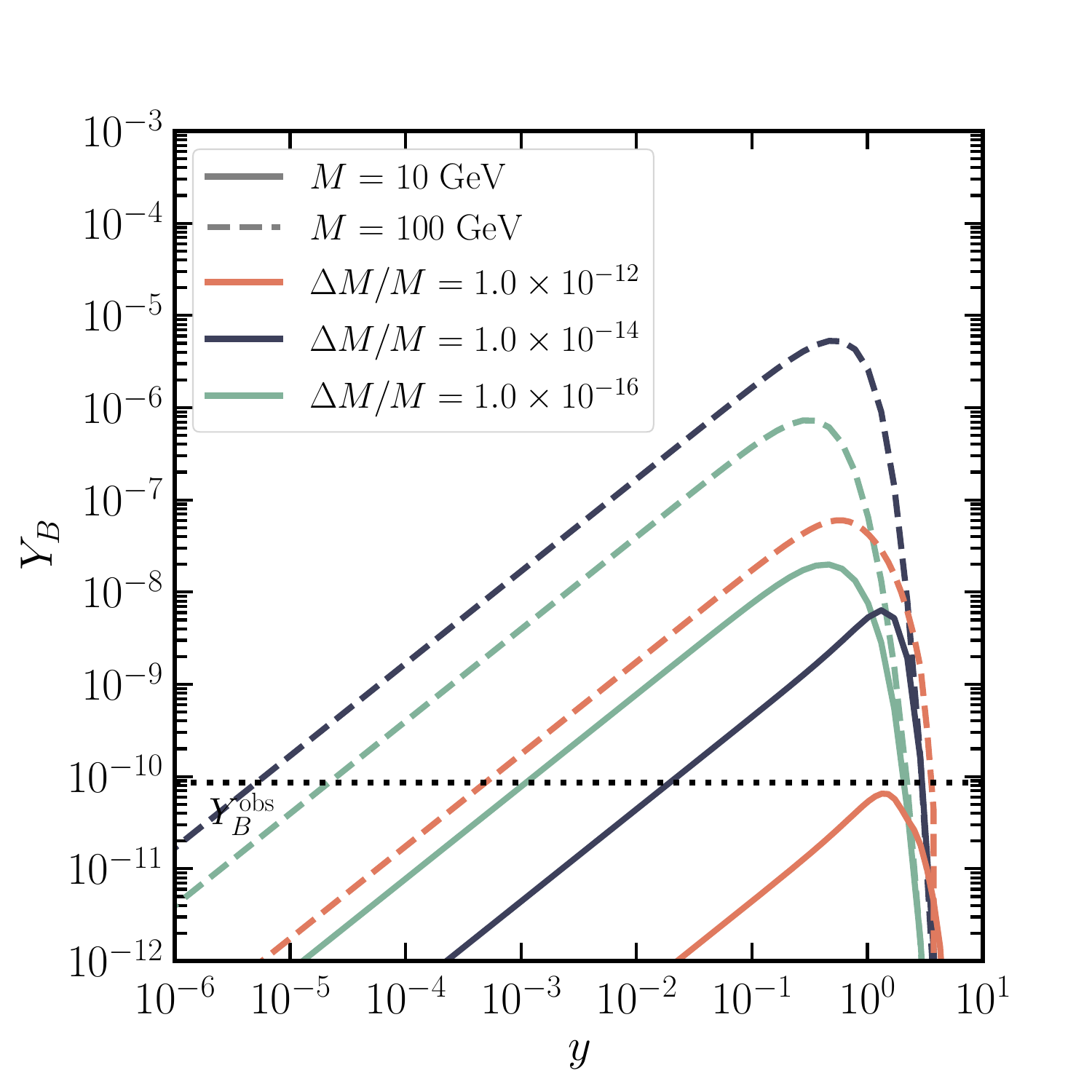}
    \caption{\label{fig:1}  The behaviour of the final baryon asymmetry yield as a function of the angle $y$ for different values of the RHN mass scale $M$ (different line styles) and relative mass splitting $\Delta M /M$ (different colors). The horizontal dotted line represents the observed baryon yield.}
\end{figure}
\begin{figure}[t!]
    \centering
    \includegraphics[width = 0.9 \linewidth]{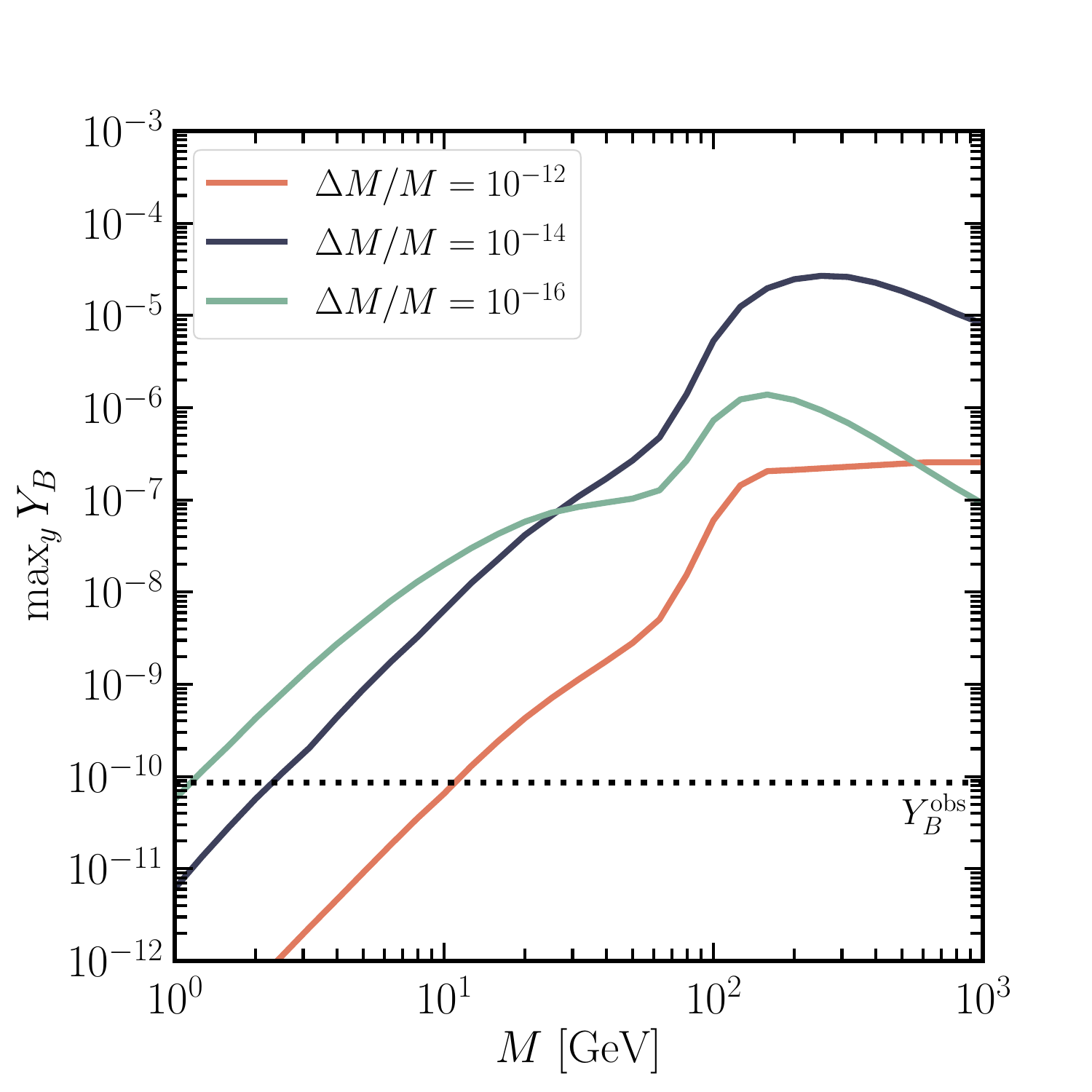}
    \caption{\label{fig:2}The maximum value of the baryon asymmetry yield (with respect to the angle $y$) as a function of the RHN mass scale $M$, for different relative mass splitting $\Delta M / M$. The horizontal dotted line represents the observed baryon yield.}
\end{figure}

In Fig.~\ref{fig:1} we show the behavior of the baryon asymmetry in some benchmark cases and we compare it with its observed value (horizontal dotted line). In particular, we consider two benchmark values for the RHN mass scale $M=10~{\rm GeV}$ (solid lines) and $M=100~{\rm GeV}$ (dashed lines), for three different values for the relative mass splitting $\Delta M / M = 10^{-12},\,10^{-14},\,10^{-16}$. In the benchmark cases we show, enough baryon asymmetry can be achieved between an upper and lower limit in $y$. These translate to upper and lower limits in $U^2$, the latter coinciding with the seesaw limit.
\begin{figure}[t!]
    \centering
    \includegraphics[width = 0.9 \linewidth]{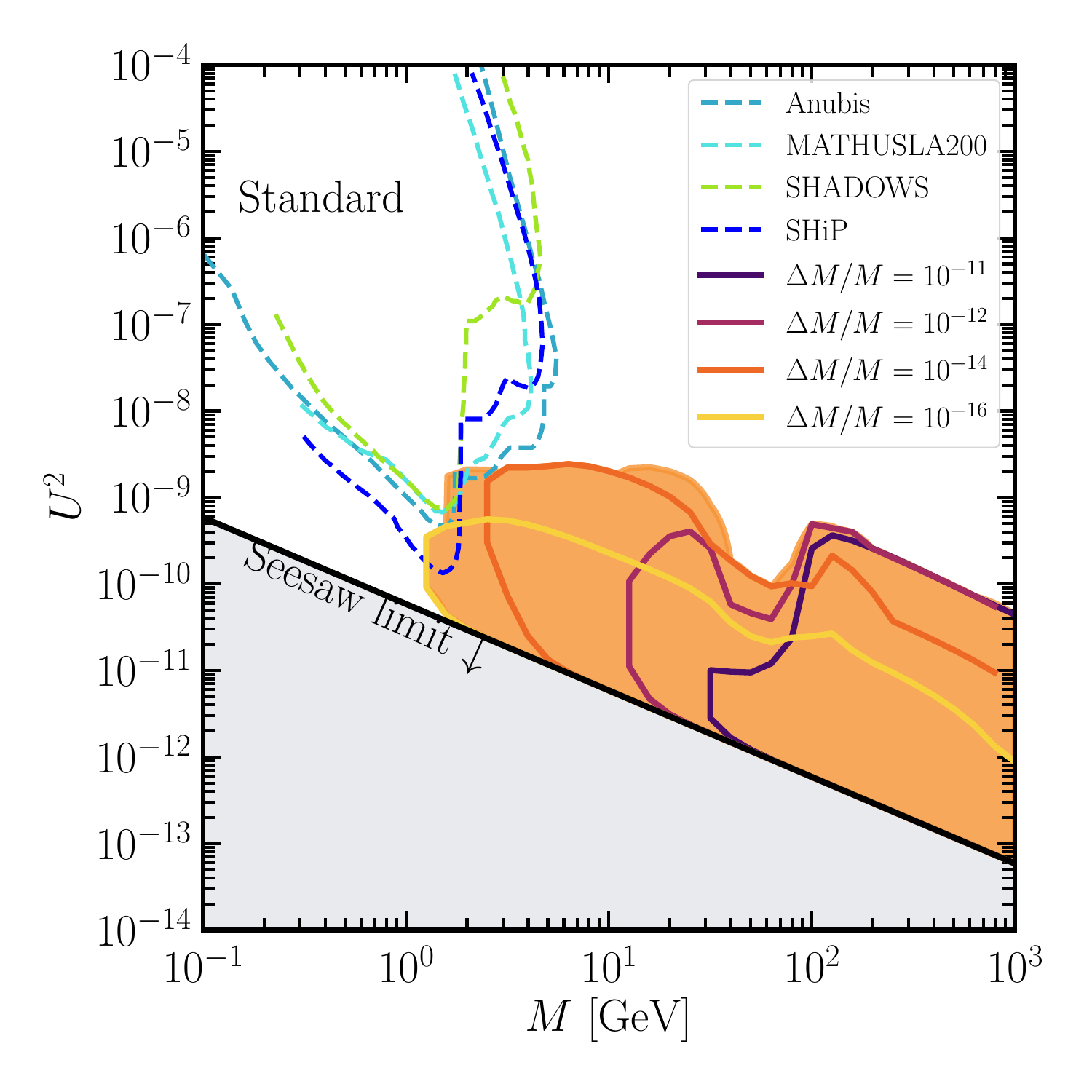}
    \caption{\label{fig:3} The allowed values for the RHN mass $M$ and the mixing parameter $U^2$ (defined in Eq.~\eqref{eq:U2}) for which the standard sub-TeV resonant leptogenesis is achieved. The different lines correspond to some benchmark values of the relative mass splitting $\Delta M / M$, whose superimposition defines the orange region. The gray area is theoretically excluded by the seesaw mechanism. We report for comparison the projected constraints from the SHiP \cite{SHiP:2018xqw}, SHADOWS \cite{Baldini:2021hfw}, MATHUSLA \cite{MATHUSLA:2022sze} and ANUBIS \cite{Hirsch:2020klk} experiments. }
\end{figure}

In Fig.~\ref{fig:2} we report the value of $Y_B$ maximised over $y$, as a function of the RHN mass scale, for different mass splittings. Hence, the requirement of successful leptogenesis also defines a lower bound on the RHN mass scale at $\mathcal{O}({\rm GeV})$, which is of high relevance for the experiments devoted to heavy neutral lepton searches.

Hence, we scan over the practically-unobservable relative mass splitting $\Delta M/M$ in the range from $10^{-16}$ to $10^{-11}$, and we obtain the region in the plane $M$-$U^2$ for which the correct baryon asymmetry is achieved. We choose this range in $\Delta M/M$ in order to maximise the available parameter space for leptogenesis. Considering larger values of the mass splitting does not change our results. Assuming instead that $N_i$ have a vanishing initial abundance  This is shown in Fig.~\ref{fig:3} by the orange region. The colored lines display different contours in fixed relative mass splitting $\Delta M/M$ for which $Y_B = Y_B^{\rm obs}$. The gray area is instead theoretically excluded by the seesaw mechanism. Our results are compatible to the ones presented in Refs.~\cite{Granelli:2020ysj, Klaric:2020phc}. We emphasize that in our treatment we use the full thermally corrected form of the decay rates as done in Ref.~\cite{Hambye:2016sby}, and we improve the calculations of the sphalerons freeze-out by taking into account the non-instantaneous case. Moreover, we underline that we do not interpolate the RHN self-energy function $\gamma$ in Eq.~\eqref{eq:I} in order to approximate the effect of soft gauge scatterings, finding the dip feature in our result for $M\lesssim 100~{\rm GeV}$ and high $U^2$ values.

\section{Non-standard cosmology from PBHs \label{sec:PBH}}

In this work we are interested in how the presence of PBHs affects the history of the Universe and, consequently, the scenario of sub-TeV resonant leptogenesis. For the sake of simplicity, we consider a population of Schwarzschild black holes with a monochromatic mass distribution. In this case, the physics of the non-rotating PBHs is described by two parameters only: their mass $\MPBH$ at formation and their abundance $\beta^\prime$, which is defined by
\begin{equation}
\label{eq:betap}
    \beta^\prime = ({\gamma_{\rm PBH}})^{\frac12}\frac{\varrho_{\rm PBH}}{\varrho_{\rm rad}}\bigg|_{\rm in} \,,
\end{equation}
where $\gamma_{\rm PBH}\simeq0.2$ is a commonly-adopted efficiently factor, and $\varrho_{\rm PBH}$ and $\varrho_{\rm rad}$ indicate the comoving energy densities of PBH and radiation, respectively. From the moment of formation until they evaporate completely, the PBHs produce every elementary particle species (including RHNs) with mass smaller than their Hawking temperature $T_{\rm PBH} \approx 10^7 (10^{6}{\rm g}/\MPBH)~{\rm GeV}$. PBHs lighter than $\sim 10^6~{\rm g}$ evaporate completely while the sphalerons are still efficient, thus providing an additional non-thermal source of RHNs. Such a scenario has been well studied in Refs.~\cite{Bernal:2022pue,Perez-Gonzalez:2020vnz}. On the other hand, PBHs heavier than $\sim 10^6~{\rm g}$ evaporate completely after the sphaleron freeze-out and the production of low mass RHNs is negligible compared to the thermal abundances, especially for $T>\Tsphal$.
\begin{figure*}[t!]
    \centering
        \includegraphics[width = 0.45\textwidth]{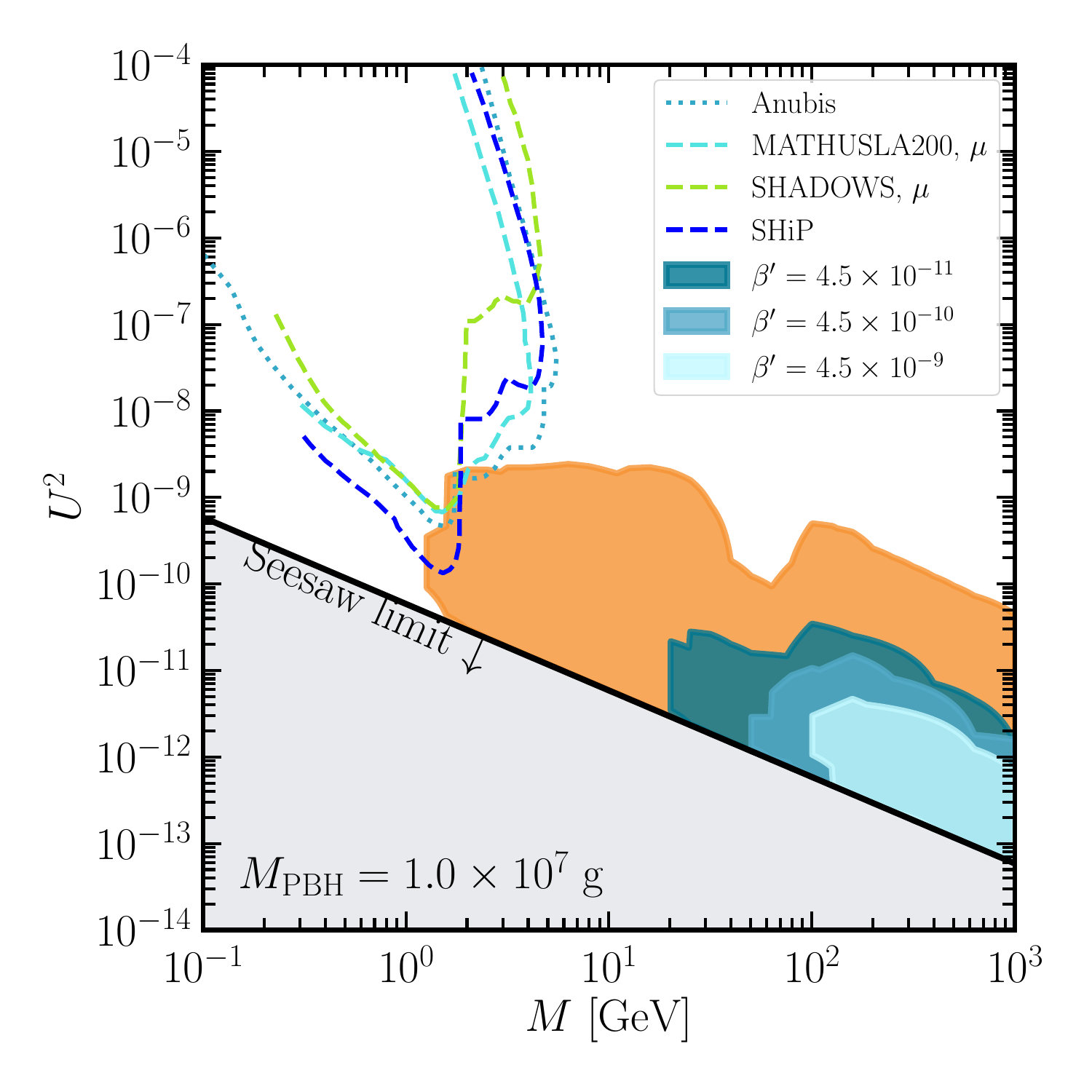}
        \hspace{0.05\textwidth}
        \includegraphics[width = 0.45\textwidth]{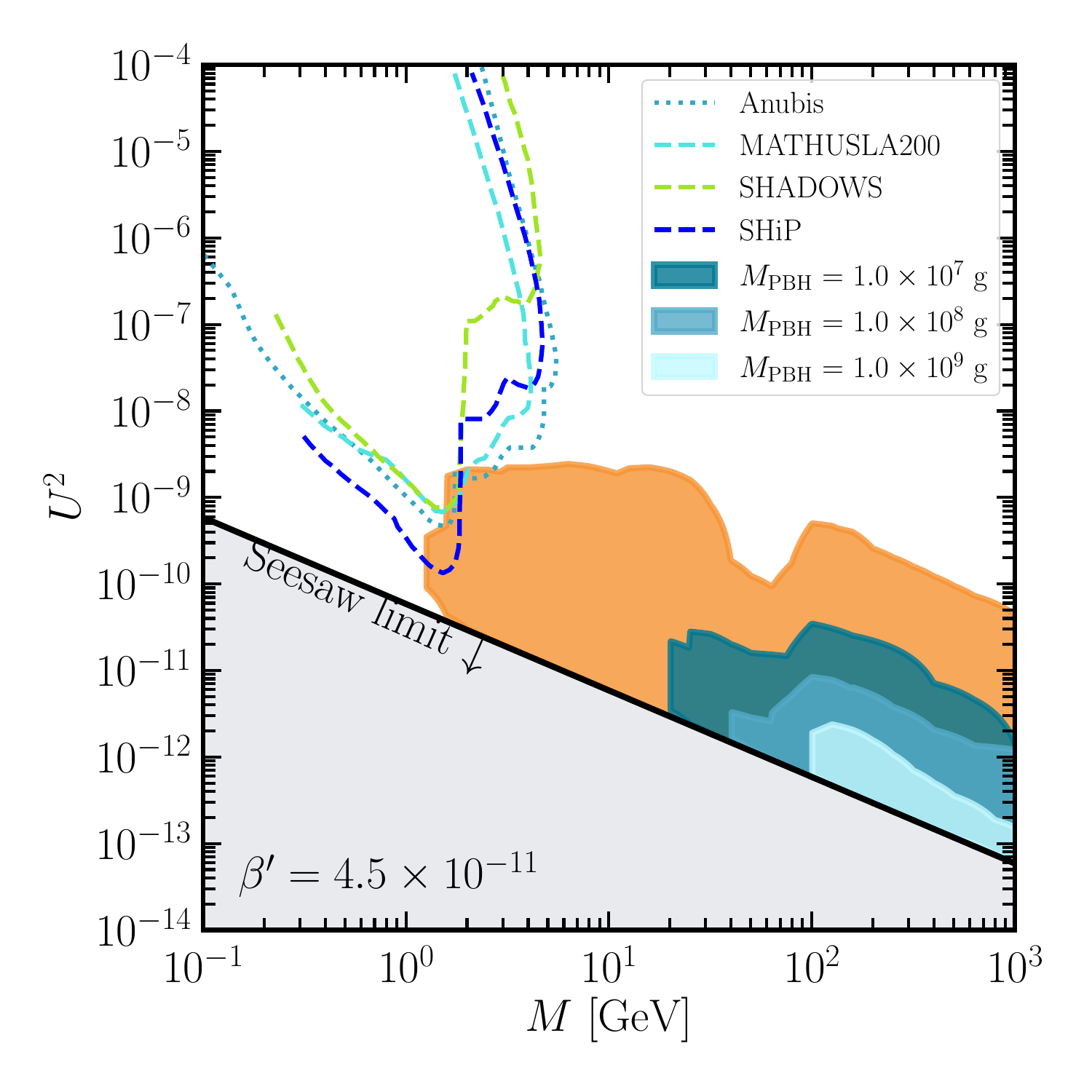}
       \caption{\label{fig:4} Allowed regions in the RHNs parameter space achieving successful sub-TeV resonant leptogenesis in case of different values for the PBHs abundance fixing $M_{\rm PBH} = 1.0\times 10^7~{\rm GeV}$ (left panel) and for the PBHs mass fixing $\beta^\prime = 4.5 \times 10^{-11}$ (right panel). The orange region is the standard result without PBHs already discussed in Fig.~\ref{fig:3}.We report for comparison the projected constraints from the SHiP \cite{SHiP:2018xqw}, SHADOWS \cite{Baldini:2021hfw}, MATHUSLA \cite{MATHUSLA:2022sze} and ANUBIS \cite{Hirsch:2020klk} experiments.}
\end{figure*}

We restrict our study to the second scenario, and consider PBHs which evaporate between the sphaleron freeze-out at $T=\Tsphal$ and the onset of Big Bang Nucleosynthesis, setting the mass range to $10^6~{\rm g} \leq \MPBH \lesssim 10^9~{\rm g}$. In this case, the presence of PBHs modifies the generation of the baryon asymmetry through two main effects. The first is the dilution of the final baryon asymmetry due to the complete evaporation of PBHs, which is associated with entropy injection described by 
\begin{equation} 
\frac{{\rm d}\mathcal{S}}{{\rm d}\alpha} = -\frac{f_{\rm SM}}{T} \frac{{\rm d}\ln \MPBH}{{\rm d}\alpha} \varrhopbh \,,
\end{equation}
where $f_{\rm SM}$ is the fraction of Hawking radiation composed of SM particles, and $\mathcal{S}$ is the comoving entropy density. The injection of entropy is significant when $\varrhopbh \geq \varrhorad$ during evaporation.

The second effect is more subtle and concerns the modification of the standard cosmological evolution due to a period in which PBHs dominate the energy content of the Universe (matter-domination epoch). This is encoded in the Hubble rate, which takes the expression
\begin{equation}
    H^2 = \frac{8\pi}{3\Mpl^2}\left(\frac{\varrhopbh}{a^3}+\frac{\varrhorad}{a^4} \right) \,.
\end{equation}
When $\varrhopbh \geq \varrhorad$, the Hubble rate becomes enhanced with respect to the standard case, and dynamically alters the evolution of the asymmetry. As shown in Ref.~\cite{Calabrese:2023key}, the dominance of PBHs also alters the sphaleron freeze-out temperature $\Tsphal$. While this has no implications in the case of high-scale thermal leptogenesis, where the asymmetry generation occurs well before the sphaleron freeze-out, it becomes pivotal in sub-TeV resonant leptogenesis. Indeed, in this case, the generation of the asymmetry proceeds until the sphaleron freeze-out, making any modification of this temperature a crucial determinant in the final amount of baryon asymmetry.

We track the evolution of the PBH masses and energy density, and the effects on radiation density, temperature, entropy and the Hubble rate by numerically solving the coupled set of equations reported in our previous paper~\cite{Calabrese:2023key}.

\section{Results \label{sec:results}}

We compute the final baryon asymmetry by solving the set of coupled Boltzmann equations including the ones related to the PBH evolution. Hence, we obtain the allowed regions achieving the right final baryon asymmetry in the combined parameter space of RHNs and PBHs, which consists of five parameters: the RHNs mass scale $M$, the RHNs mass splitting $\Delta M$, the real angle $y$ in the Casas-Ibarra matrix, the PBH mass scale $M_{\rm PBH}$, and the PBHs initial abundance $\beta^\prime$.

In Fig.~\ref{fig:4} we show how the RHNs parameter space providing successful leptogenesis is modified due to the presence of PBHs and their evaporation, for different PBHs abundance (left panel) and different PBHs mass (right panel). In the plots, the regions are obtained by superimposing different values for the RHNs mass splitting $\Delta M$ as done in Fig.~\ref{fig:3}. Hence, the effect of PBHs is to shrink the allowed region for the RHNs parameters towards higher masses $M$ and smaller mixing parameters $U^2$. Indeed, the entropy injection from PBHs evaporation reduces the final yield $Y_B$ of baryon asymmetry and tightens the allowed range for the parameters $M$ and $y$ (see Fig.~\ref{fig:1}) in a non-linear way, due to the
non-trivial effects on the sphaleron freeze-out and the
Hubble rate. We have also reported in Fig.~\ref{fig:4} the projected sensitivity curves for the planned SHiP \cite{SHiP:2018xqw}, SHADOWS \cite{Baldini:2021hfw},MATHUSLA \cite{MATHUSLA:2022sze} and ANUBIS \cite{Hirsch:2020klk} experiments. While these experiments can probe the very edges of resonant leptogenesis parameter space, much larger regions would be ruled out by even the tiny populations of PBHs as illustrated in the figure. Therefore the strong tension between PBHs and leptogenesis can shed light on parameter space well out of reach of direct detection experiments.

\begin{figure}[t!]
    \centering
        \includegraphics[height= 0.9\linewidth]{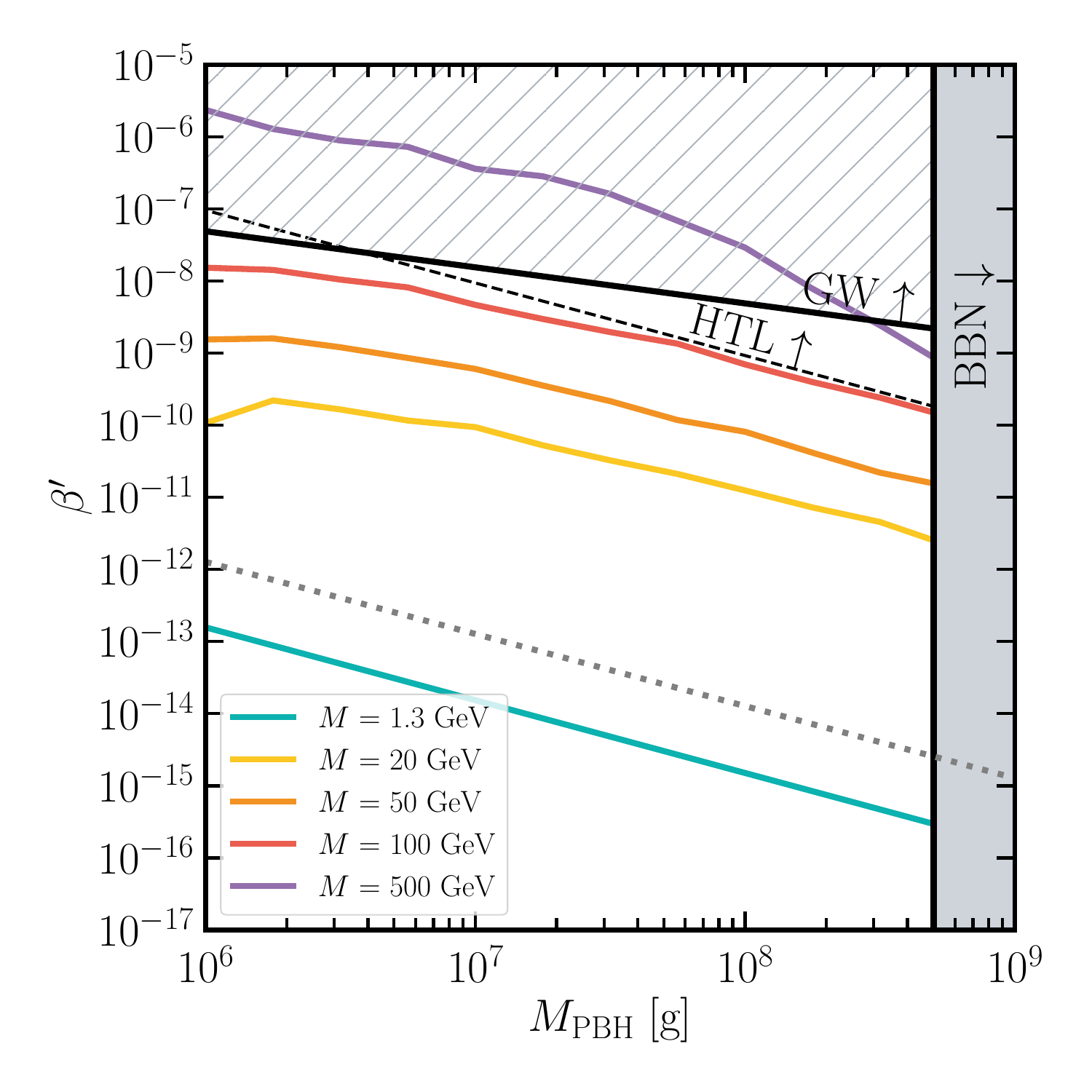}\,\,
       \caption{\label{fig:5} Upper bounds on the PBHs abundance $\beta^\prime$ defined in Eq.~\eqref{eq:betap} as a function of the PBH mass $M_{\rm PBH}$ for different values of the RHNs mass scale, potentially detectable by upcoming experiments. The hatched region is excluded by the constraints from GW energy density from BBN observations, while the gray one from the PBH evaporation during BBN. For comparison, the dashed line bounds from below the region excluded by the scenario of High-scale Thermal Leptogenesis discussed in Ref.~\cite{Calabrese:2023key}. The gray dotted line bounds from below the region where we observe a period dominated by PBHs in the thermal history of the Universe.}
\end{figure}

Our findings also have interesting implications on the PBHs parameter space. In Fig.~\ref{fig:5} we report the constraints on the PBHs abundance as a function of the PBHs for different RHNs mass scales. These constraints delineate the regions of PBHs parameter space which are incompatible with sub-TeV resonant leptogenesis at different mass scales M. The smaller the RHNs mass scale, the stronger the constraints on $\beta^\prime$. Indeed, for $M\sim\mathcal{O}({\rm GeV})$ the maximum allowed baryon asymmetry that can be produced by the sub-TeV resonant leptogenesis is of the same order of the one experimentally observed. In the same plot, we show the limits from the observations of Big Bang Nucleosynthesis (shaded gray region) and from the constraints (hatched region) on the energy density of primordial Gravitational Waves~\cite{Papanikolaou:2020qtd,Domenech:2020ssp,Papanikolaou:2022chm}. Moreover, we also report the constrains we placed in our previous study based on the assumption that the baryon asymmetry of the universe is produced through the High-scale Thermal Leptogenesis (HTL)~\cite{Calabrese:2023key}. Hence, the intersection between the regions above the HTL dashed black line and the one above the solid purple line can be regards as the most conservative constraints on the PBHs parameter space placed by the scenarios of baryogenesis via leptogenesis.

\section{Conclusions \label{sec:conclusions}}

In the present paper, we have investigated the impact of PBHs with masses from $10^6$ to $\sim10^9$~g on the predictions of sub-TeV resonant leptogenesis. Differently from the standard scenario of high-scale leptogenesis, in the framework of sub-TeV resonant leptogenesis the observed baryon asymmetry of the Universe is attained through the interactions of (at least) two almost-degenerate Right-Handed Neutrinos with mass from 1~GeV to few TeVs. For this reason, such a scenario is of great interest since it can be probed in laboratories dedicated to the search of heavy neutral leptons through their mixing with active neutrinos.

In the standard cosmological evolution without PBHs, we have solved the Boltzmann equations for the number densities of the Right-Handed Neutrinos and the baryon asymmetry accounting for the non-instantaneous sphaleron freeze-out. In particular, we have assumed the Right-Handed Neutrinos to be initially in thermal equilibrium with the photon bath implying that neutrino oscillations do not play a crucial role in determining the final baryon asymmetry. Hence, we have obtained the standard result (see Fig.~\ref{fig:3}) for the allowed region in the neutrino parameter space which mainly consists of two parameters: the mass scale $M$ of the two Right-Handed Neutrinos and the quantity $U^2$ parameterising the active-sterile neutrino mixing. We have marginalized over the Right-Handed Neutrino mass splitting $\Delta M$.

Then, we have studied the evolution of PBHs whose effects are the modification of the Hubble rate through a matter-domination epoch and the injection of entropy from their evaporation occurring after the sphaleron freeze-out. Interestingly, we have found that the allowed neutrino parameter space shrinks towards higher Right-Handed Neutrino mass scales and smaller active-sterile mixing parameter for different masses and initial abundances of PBHs (see Fig.~\ref{fig:4}). Hence, we have pointed out that the potential laboratory detection of Right-Handed Neutrinos in the mass range from 1 to $\sim 100$~GeV would allowed us to place very competitive constraints on the abundance of PBHs (see Fig.~\ref{fig:5}).

\section*{Acknowledgments}

This work was partially supported by the research grant number 2022E2J4RK ``PANTHEON: Perspectives in Astroparticle and Neutrino THEory with Old and New messengers'' under the program PRIN 2022 funded by the Italian Ministero dell'Universit\`a e della Ricerca (MUR) and by the research project TAsP (Theoretical Astroparticle Physics) funded by the Istituto Nazionale di Fisica Nucleare (INFN).

\bibliographystyle{apsrev4-1}
\bibliography{Bibliography}

\newpage

\end{document}